# Démarche d'évaluation de l'usage et des répercussions psychosociales d'un environnement STIC sur une population de personnes âgées en résidence médicalisée.

*Christine Michel[1], Marc-Eric Bobillier-Chaumon[2], Véronique Cohen Montandreau[2], Franck Tarpin-Bernard[1, 3]*

[1] ICTT
Bat L. de Vinci - INSA de Lyon
21, avenue Jean Capelle
69621 VILLEURBANNE
Cedex, France

[2] ICTT
Ecole Centrale de Lyon
36, av. Guy de Collongue
69134 ECULLY Cedex,
France

[3] SBT
Bâtiment CEI, 66 Bd Niels
Bohr - BP 2132
69603 VILLEURBANNE
Cedex, France

Christine.Michel@insa-lyon.fr, Marc-Eric.Bobillier@ec-lyon.fr, montandreau@iuta.univ-lyon1.fr
Franck.Tarpin-Bernard@insa-lyon.fr

**RESUME**
Le projet MNESIS a pour objectif de voir si l'utilisation d'environnement STIC par des personnes âgées en résidences médicalisées stimule leurs capacités cognitives et contribue à leur meilleure intégration, reconnaissance ou acceptation au sein de leur environnement social, familial et médical. Dans cet article nous présentons la démarche d'évaluation définie pour vérifier cette hypothèse. A la frontière des protocoles classiques centrés utilisateurs (enquêtes et observation indirecte) et des études de Web Usage Mining (construction de bases de connaissances sur les usages à partir des traces), elle se base sur un mode de collecte d'informations, de manière directe et indirecte, à grande échelle et sur de longues périodes ; une organisation qui les rend éventuellement réutilisable (construction d'une base de connaissance sur les usages) et une interrogation personnalisée.

**MOTS CLES :** évaluation, interface, usage, intégration sociale, modélisation

**ABSTRACT**
The MNESIS Project aims to see whether the use of computerized environment by elderly people in medicalized residences stimulates their cognitive capacities and contributes to a better integration, recognition or acceptance within their social environment (friends, family, medical staff). In this paper we present the protocol of evaluation that is defined to check this assumption. This protocol is between users' centred traditional protocols (built on investigations and indirect observation) and studies of Web Usage Mining (where knowledge databases about the uses are built from traces of use). It allows collecting direct and indirect information on a large scale and over long periods.

**CATEGORIES AND SUBJECT DESCRIPTORS:** H.5.2 [User Interfaces] : *Evaluation/methodology*, H.5.1 [Multimedia Information Systems] : *Evaluation/ methodology* , H.1.2 [User/Machine Systems] : *Human factors,* D.2.5 [Testing and Debugging] : *Tracing*, K.4 [Computers and society] : Miscellaneous

**GENERAL TERMS:** Information Systems, Software

**KEYWORDS:** evaluation, interface, use, social integration, modeling

## INTRODUCTION

Les adultes âgés en bonne santé montrent un déclin naturel lié au vieillissement de leurs capacités cognitives et en particulier de la mémoire,. Diverses expériences ont montré que des stimulations permettent d'enrayer cela et d'améliorer les performances des sujets âgés sans altération cognitive [11]. Plus indirectement, le maintien des capacités cognitives aurait un impact sur le degré d'autonomie dans les tâches de la vie quotidienne. [10]. Dans ce contexte général, le projet MNESIS propose d'étudier les répercussions cognitives et psychosociales liées à l'usage d'un dispositif STIC sur ce type de population ainsi que les effets induits par ces usages sur les relations que ces personnes tissent avec leur environnement social et familial. Pour ce faire, nous avons choisi d'observer directement, sur une grande échelle et de manière quantitative, l'usage effectué et, indirectement, de manière qualitative, les incidences psychosociales. Le croisement de ces méthodes contribue à l'objectivation des résultats tant d'un point scientifique que d'un point de vue écologique (significativité des données recueillies). Comme le laissait entendre Vermersch [6], les observables sont une source d'information très précieuse à la fois directement pour ce dont ils témoignent et indirectement dans la mesure où ils corroborent les informations qui pourront être abordées par les verbalisations.

**LE PROTOCOLE EXPERIMENTAL**

Les protocoles d'évaluation centrés utilisateur que l'on trouve classiquement en IHM s'organisent généralement de la manière suivante : les interactions possibles entre l'homme et l'interface du système sont renseignées selon différents modèles qui sont confrontés. L'utilisateur en particulier est modélisé de manière théorique, son comportement est renseigné à partir de données recueillies au moyen de questionnaires, entretiens, enregistrements vidéo qui s'obtiennent soit en contexte, soit dans des laboratoires d'utilisabilité [2]. Parfois, comme c'est le cas dans le laboratoire Multicom [5], des traces d'usage sous forme de capture d'événements souris-clavier et de vidéo de l'interface sont utilisées mais leur exploitation et analyse se fait de manière manuelle comme aide contextuelle à l'interprétation faite par les ergonomes ou par des traitements statistiques simples (comptage, moyenne, …). Une autre approche [9] présente un protocole où les données d'usage enregistrées combinent des enregistrements souris-clavier et les mouvements oculaires capturés avec un dispositif d'eye tracking. Là non plus, aucune exploitation poussée n'est faite de ces données analysées selon des traitements statistiques simples. Ce type de méthode ne peut être utilisé dans notre contexte qui exige : (1) qu'on ne perturbe pas les utilisateurs (en les sortant de leur contexte de vie pour les emmener dans des laboratoires d'utilisabilité ou qu'on leur demande d'utiliser un casque d'eye tracking par exemple), (2) qu'on ne les bride pas dans leurs initiatives (en ne leur permettant l'accès au système que pendant des périodes bien définies), (3) qu'on observe concrètement quand et comment ils utilisent le système de manière à saisir la réalité de pratiques sociales à grande échelle de façon précise et non subjective, en particulier en ce qui concerne le temps et les actions réalisées, (4) que l'on vérifie la réalité de leur représentation du système et des usages effectués de manière à comprendre le processus d'acceptation et d'appropriation de cette technologie.

Pour percevoir et modéliser l'utilisation du système, d'une manière fine et sans perturber l'utilisateur, nous avons utilisé les méthodes développées pour la conception d'hypermédias adaptatifs. Les données analysées sont généralement : d*es données environnementales* permettant l'adaptation de l'interface et d*es données d'usage*, permettant l'adaptation du comportement du système. Ces dernières sont significatives des interactions de l'utilisateur avec le système et sont mises en évidence par des études de *profiling* et de *tracking*. Ces données peuvent aussi être utilisées a posteriori comme dans les études de Web Usage Mining [8], pour mettre en évidence des comportement généraux d'usage. Les études ont d'abord consisté à traiter les fichiers de trace (*log*) pour en extraire des informations utilisables, et à leur appliquer des techniques mathématiques robustes pour en extraire du sens [1] (regroupement, classification, régularité de comportements par exemple). Ce type de traitement a rapidement trouvé ses limites, les données n'étant réellement exploitables que si elles étaient structurées. Ainsi, de réelles bases de connaissance sur les usages ont été développées. Ce type d'organisation a été choisi pour structurer l'ensemble des données de notre évaluation. Le modèle générique choisi est MUSETTE (Modéliser les USages Et les Tâches pour Tracer l'Expérience) [4]. Il a été conçu à l'origine pour produire des conteneurs de connaissance exploitables dans le cadre d'assistants automatiques de l'utilisateur dans la réalisation d'une tâche.

Pour identifier l'impact d'un dispositif STIC sur des personnes âgées nous étudions, sur une période de 6 mois, le comportement de résidents de différentes maisons de retraite participant à des activités diverses stimulant leurs capacités cognitives, soit par le biais d'un logiciel, *Activital*$^{TM}$ [7], soit par le biais d'activités équivalentes mais réalisées sans dispositif technique. Développé par la société SBT pour un public spécifique de seniors [3], il propose trois activités complémentaires : un ensemble de *jeux cognitifs*, un outil de *rédaction de journal* de résidence pour développer la créativité et un outil de *messagerie électronique* simplifié pour favoriser les liens sociaux et la communication. Trois groupes de résidents ont été constitués. Le groupe I dispose d'un équipement informatique et utilise lors de séances encadrées par un animateur les fonctions de jeu et messagerie. Le groupe II a la même configuration technique mais utilise les fonctions de messagerie et journal. Le groupe III ne dispose d'aucun dispositif informatique et réalise avec l'animateur des ateliers d'écriture ou de jeux dont le contenu est équivalent à ceux d'*Activital*$^{TM}$. Chaque groupe est composé de 45 résidents environ, pratiquant les activités à raison de 2 séances par semaine. La comparaison des résultats de l'évolution cognitive (réalisés par le laboratoire EMC, Université Lyon II), de l'intégration sociale et de l'image de soi, des résidents des groupes I et II avec le groupe III permet de définir l'impact sur ces points du logiciel par rapport à une stimulation plus classique dans ces contextes (la comparaison des groupes I et II montre plus précisément quelles activités du logiciel pourraient être les plus marquantes). L'observation des traces d'usage et des résultats des entretiens sociaux permet de déterminer le mode de prise en main du logiciel et les autres répercussions éventuelles. Commencée en décembre 2004, l'expérimentation en est actuellement à mi-parcours. La présentation de la collecte et de l'organisation des données nous permettra d'expliciter plus précisément comment nous allons le faire.

**LES DONNEES D'ANALYSE**

La collecte des données est effectuée manuellement pour les données sociales (sous la forme d'entretiens ou d'enquêtes) et automatiquement pour les données d'usage (des interactions spécifiques de l'utilisateur avec l'interface sont enregistrées tout au long de l'expérimentation). Ces données sont structurées selon le modèle MUSETTE introduit précédemment, il permet de modéliser, à partir des interactions, des tâches particulières. La combinaison ou succession de ces dernières avec les données sociales, permet de mettre évidence des profils d'usages spécifiques. Les paragraphes suivant décrivent successivement les données collectées et les profils d'usages significatifs identifiés ; la présentation globale du mode d'organisation et de traitement des données étant relativement longue et spécifique, elle sera faite dans une autre publication.

*Données d'usage :* On retrouvera des événements classiques comme le survol ou le clic d'un bouton, ou encore le déplacement ou le redimensionnement d'un objet. D'autres événements sont spécifiques à l'application comme les scores aux jeux ; ou encore calculés en fonction du contexte comme le bloc de texte, donnée quantifiée de l'activité textuelle (rédaction de mail, rédaction d'article) qui ne prend pas en compte le contenu textuel de manière à préserver l'intimité des utilisateurs. A ces évènements s'ajoutent des variables contextuelles comme l'identifiant de l'utilisateur, la date, l'activité concernée ou encore les données sociales. Ces interactions élémentaires sont utilisées pour décrire les tâches représentant le cœur de l'activité.

*Données sociales :* Les techniques proposées ont toutes pour but d'apprécier les incidences effectives des artefacts techniques sur l'évolution de la structure sociale, familiale et médicale du résident et plus généralement sur la recomposition de ses liens et ses pratiques sociales. Afin de confronter les perceptions des différents acteurs susceptibles d'être touchés (de manière directe ou indirecte) par le dispositif technologique, le recueil de données porte à la fois sur les usagers du logiciel et sur leurs entourages : familial, social et médical. Il renseigne les caractéristiques sociales, biographiques, familiales, etc. de la population des utilisateurs et, d'autre part, recense des variables agissant sur le "confort social" et sur l'établissement des relations sociales.

La collecte des données sociales repose sur différentes techniques et outils de la psychosociologie. Elles permettent de rendre compte de différentes dimensions psychosociales des résidents (liens, pratiques, intégration, estime de soi) et d'appréhender la construction du lien social. Elles sont présentées ci-après.

**Définition du profil psychosocial du résident**. L'objectif est de disposer de premiers éléments d'informations sur le résident et de créer une relation de confiance incitatrice, pour des personnes en situation de handicap (mobilité réduite, organes sensorielles altérées) et souvent isolées socialement, à se livrer spontanément dans le cadre d'une investigation plus personnelle.

**Mesure de l'estime de soi du résident.** L'objectif est d'évaluer la manière dont les individus se perçoivent, la confiance qu'ils ont en eux et s'ils ont l'impression de pouvoir réaliser leurs aspirations. La méthode utilisée est une échelle de mesure d'estime de soi.

**Analyse des pratiques sociales et du lien social des résidents.** L'objectif est d'appréhender la "dynamique sociale" des résidents, c'est-à-dire la nature, la quantité, la qualité et la fréquence de leurs relations sociales, d'apprécier la réalité de ces liens sociaux en observant concrètement leurs pratiques sociales et d'évaluer la perception qu'a leur entourage sur ces diverses pratiques et interactions sociales. Les méthodes utilisées sont le questionnaire, le sociogramme (qui permet d'estimer la place et le positionnement social des individus dans un groupe les uns par rapport aux autres, en termes de personnes attractives, invisibles, "répulsives"…) et l'observation par carnet de bord.

**Description des attentes et des premières réactions et usages face à l'outil.** L'objectif est d'appréhender les représentations des résidents et de leurs entourages sur les technologies proposées et d'analyser les premières impressions de manipulation de l'outil ainsi que l'appropriation qui peut en être faite. Les méthodes utilisées sont le questionnaire et l'observation des séances d'activité (par grille d'observation de la médiation entre l'animateur et les résidents : nous relevons le type d'assistance demandée – technique, opératoire, psychologique… –, pour quels types de tâches et de technologie cette assistance est sollicitée et leurs occurrences sur une séance de formation).

## CARACTERISTIQUES GENERALES DES USAGES ET DES USAGERS

*Les profils d'usager* Les profils d'usagers sont définis conjointement à partir des enquêtes et à partir des traces. Les enquêtes permettent de définir les *tempéraments généraux* des utilisateurs et leurs *comportement face à l'implantation du nouveau système* (« acquis », « transfuges déçus », « transfuges convertis », « réfractaire par expérience », « réfractaires traditionalistes »). Les traces permettent de définir des *profils d'usage global* du système (utilisateur « curieux », « peureux », « complet ») ou spécifiques à l'une des 3 applications. Pour chacune d'elle nous mettons en évidence respectivement les *profils communicationnels* (« individuel », « sociable », « attractifs » ou « invisibles ») à partir de l'utilisation et la variété des contacts de la messagerie ; les *profils joueurs* (« zappeur », « consciencieux », « dirigé »), et les *profils rédacteurs* (« fouilleur », « transcripteur », « e-redacteur »).

*Acquisition d'une culture technique pour l'utilisateur* La mise en évidence d'un apprentissage dans l'*utilisation du logiciel et de l'environnement informatique* est visible par une évolution de la manipulation des services proposés, une amélioration de la dextérité et un gain d'autonomie. Si un apprentissage est observé nous déterminons *le type d'apprentissage* en particulier si nous sommes en présence d'une intégration verticale (utilisation de séquences d'action identifiées) ou dans une logique transversale (affordance, digression d'usage positif ou négatif, changement de niveau). L'*acceptation des technologies* est identifiable par les critères d'utilité, d'utilisabilité et d'accessibilité. L'*amélioration psychomotricielle* est quantifiable par la rapidité d'interaction de l'individu avec l'environnement. Enfin, nous observons si le logiciel joue le rôle de *« facilitateur technologique »*, c'est à dire qu'il stimule l'usager dans l'utilisation d'autres technologies (téléphone portable, appareils numériques). Certaines caractéristiques sont clairement identifiables uniquement à partir des traces comme l'*utilisation du logiciel et de l'environnement informatique* ou *le type d'apprentissage.* D'autres, comme l'utilisation de technologies externes au système, ne s'interprètent qu'à partir des réponses des utilisateurs à l'enquête.

**L'*intégration sociale*.** Pour comprendre l'incidence du dispositif sur l'intégration sociale nous allons observer s'il y a modification de la *définition de soi* et en particulier : comment l'utilisateur se compare et se définit par rapport aux autres et ainsi se situe dans le champ social, si l'estime qu'il a de lui-même (indiquée dans les questionnaires) est cohérente ou non avec ses performances (score, dextérité, …) d'utilisation du système et enfin si l'utilisation a renforcé sa tendance naturelle ou l'a fait évoluer. Nous observerons ensuite s'il y a une *modification de la reconnaissance sociale* et *du lien social*, identifiable en particulier par l'évolution communicationnelle et d'action de l'utilisateur. Enfin nous analyserons le *degré d'appropriation de l'outil technologique des résidents* hors des phases d'entraînement.

## CONCLUSION

Le protocole d'évaluation propose une démarche pour combiner des données concrètes d'usage et des données d'enquête de manière à représenter l'usage réel et l'impact de l'utilisation d'un logiciel STIC pour un public de personnes âgées. La collecte normalisée dans une base de connaissance sur les usages assure une organisation des données qui les rend concrètement exploitables mais aussi réutilisables dans d'autres contextes. Plus spécifiquement, pour les évaluations en IHM centrées vers la conception et les systèmes, la modélisation de l'utilisation, en plus de la tâche ou de l'utilisateur, peut permettre de mettre en évidence et expliciter assez rapidement des phénomènes de catachrèse ou de détournement d'usage, visible actuellement uniquement sur du long terme. Le développement de cette base n'est actuellement pas finalisé, si le mode de collecte est effectivement réalisé, le mode d'interrogation et le choix des critères de modélisation des profils est actuellement en cours d'élaboration.